\documentclass[conference, peerrievew]{IEEEtran}
\IEEEoverridecommandlockouts

\usepackage[bookmarks=false]{hyperref}
\usepackage{amsmath}
\usepackage{enumitem,kantlipsum}
\usepackage{graphicx}
\usepackage[table]{xcolor}
\usepackage{booktabs}
\usepackage{threeparttable}
\usepackage{tabulary}
\usepackage{multirow}
\usepackage{units}
\usepackage[binary-units=true]{siunitx}
\usepackage{comment}

\def\BibTeX{{\rm B\kern-.05em{\sc i\kern-.025em b}\kern-.08em
    T\kern-.1667em\lower.7ex\hbox{E}\kern-.125emX}}
 
\usepackage[compact]{titlesec}
\titlespacing{\section}{0pt}{*1}{*0}
\titlespacing{\subsection}{0pt}{*1}{*0}
\titlespacing{\subsubsection}{0pt}{*0}{*0}
\usepackage[left=1.5cm,top=1.35cm,right=1.5cm,bottom=1.6cm,nofoot]{geometry}

\makeatletter
 \let\old@ps@headings\ps@headings
 \let\old@ps@IEEEtitlepagestyle\ps@IEEEtitlepagestyle
 \def\confheader#1{%
 \def\ps@headings{%
 \old@ps@headings%
 \def\@oddhead{\strut\hfill#1\hfill\strut}%
 \def\@evenhead{\strut\hfill#1\hfill\strut}%
 }%
 \def\ps@IEEEtitlepagestyle{%
 \old@ps@IEEEtitlepagestyle%
 \def\@oddhead{\strut\hfill#1\hfill\strut}%
 \def\@evenhead{\strut\hfill#1\hfill\strut}%
 }%
 \ps@headings%
 }
 \makeatother

\confheader{Published as a conference paper at the IEEE 2021 ISCAS Conference (\url{https://doi.org/10.1109/ISCAS51556.2021.9401282})
}

\begin{document}

\title{Robust and Energy-efficient PPG-based Heart-Rate Monitoring\thanks{This work was supported in part by the European
H2020 FET Project OPRECOMP under Grant 732631.}}

\author{\IEEEauthorblockN{Matteo Risso\IEEEauthorrefmark{2}, Alessio Burrello\IEEEauthorrefmark{1}, Daniele Jahier Pagliari\IEEEauthorrefmark{2}, Simone Benatti\IEEEauthorrefmark{1},\\ Enrico Macii\IEEEauthorrefmark{3}, Luca Benini\IEEEauthorrefmark{1}, Massimo Poncino\IEEEauthorrefmark{2}}
\IEEEauthorblockA{\IEEEauthorrefmark{1}Department of Electrical, Electronic and Information Engineering, University of Bologna, 40136 Bologna, Italy\\
\IEEEauthorrefmark{2}Department of Control and Computer Engineering, Politecnico di Torino, Turin, Italy\\
\IEEEauthorrefmark{3}Inter-university Department of Regional and Urban Studies and Planning, Politecnico di Torino, Turin, Italy\\
}
\IEEEauthorblockA{Corresponding Email: alessio.burrello@unibo.it}
\vspace{-1.1cm}
}
\maketitle
\begin{abstract}
A wrist-worn PPG sensor coupled with a lightweight algorithm can run on a MCU to enable non-invasive and comfortable monitoring, but ensuring robust PPG-based heart-rate monitoring in the presence of motion artifacts is still an open challenge. Recent state-of-the-art algorithms combine PPG and inertial signals to mitigate the effect of motion artifacts. However, these approaches suffer from limited generality. Moreover, their deployment on MCU-based edge nodes has not been investigated. In this work, we tackle both the aforementioned problems by proposing the use of hardware-friendly Temporal Convolutional Networks (TCN) for PPG-based heart estimation. Starting from a single ``seed'' TCN, we leverage an automatic Neural Architecture Search (NAS) approach to derive a rich family of models. Among them, we obtain a TCN that outperforms the previous state-of-the-art on the largest PPG dataset available (PPGDalia), achieving a Mean Absolute Error (MAE) of just 3.84 Beats Per Minute (BPM). Furthermore, we tested also a set of smaller yet still accurate (MAE of 5.64 - 6.29 BPM) networks that can be deployed on a commercial MCU (STM32L4) which require as few as 5k parameters and reach a latency of \unit[17.1]{ms} consuming just \unit[0.21]{mJ} per inference.
\end{abstract}



\section{Introduction}
\label{sec:intro}
Wrist-worn devices  equipped with sensors, such as wristbands and smartwatches, enable a comfortable monitoring of vital signs, hence they are becoming increasingly popular in personalized health care and medical IoT applications~\cite{yeole2016use}. 
Heart rate (HR) is one of the most critical indexes to monitor, both for activity tracking and for clinical purposes.
First generation HR-monitoring devices were based on a simple 1-3 leads ECG, connected through a chest strip, which, however, is uncomfortable or even impossible to wear in certain conditions. 
Recently, the optimization and miniaturization of photoplethysmogram (PPG) sensors has allowed to integrate HR and blood oxygenation (SPO2) monitoring in smaller, less invasive and cheaper devices~\cite{troika2014}. 
A PPG sensor consists of one or more LEDs that continuously emit light to the skin and a photodiode that measures variations of light intensity caused by blood flow, which depends on the heart rate.
A major limitation of PPG based HR estimation is represented by motion artifacts (MA) caused by variations of sensor pressure on the skin or ambient light leaking in the gap between the photodiode and the wrist. Moreover, blood flow can vary considerably depending on the type of physical activity, contributing to a less precise light absorption measurement, and hence HR estimate~\cite{DeepPPG2019}. 

Several approaches have been proposed in literature~\cite{joss2015,huang2020robust} to tackle these limitations, by
cancelling/reducing the noise caused by MAs on the PPG signal before using it to compute the HR of subjects. 
However, these methods require an extensive hand-tuning of parameters for the target dataset, leading to difficulties in generalizing over many subjects and over different activities.
Until now, limited attention has been given to deep learning approaches, despite the promising generalization results shown in \cite{DeepPPG2019,cornet2019}. 
Moreover, none of the state-of-the-art algorithms (neither classic or deep learning ones) have been yet deployed on a MCU of the class found in wrist-worn edge devices.

In this paper, we propose a collection of TCNs for HR estimation based on raw PPG and acceleration data, called \emph{TimePPG}. All TCNs are derived automatically from a single seed architecture using a NAS tool~\cite{gordon2018morphnet}, and form a Pareto frontier in the accuracy vs complexity space, from which designers can select a model based on the available computing resources. In particular, we analyze in detail three TCNs from the TimePPG family. 
The best performing model, \textit{BestMAE}, achieves a MAE of \unit[5.30]{BPM} on the popular PPGDalia dataset, and includes $\approx$ 232k trainable parameters. Coupling this TCN with simple post-processing and fine-tuning steps, we further reduce the MAE to \unit[3.84]{BPM}, outperforming the current (more complex) state-of-the-art algorithms \cite{huang2020robust}.
At the other extreme, the smallest model in TimePPG, \textit{BestSize}, uses only 5k parameters while still reaching an acceptable MAE of \unit[6.29]{BPM}. Finally, as a compromise between the former two, we analyze \textit{BestMCU}, i.e. the largest network that fits the memory of a popular MCU by STM, the STM32L476, which achieves a MAE of \unit[5.64]{BPM} with 41.7k parameters.
When deployed on the MCU, \textit{BestMCU} consumes \unit[5.17]{mJ} per inference, with a latency of \unit[427]{ms}. 
\textit{BestSize} reduces both metrics by 25$\times$, reducing energy to \unit[0.21]{mJ} and latency to \unit[17.1]{ms}.

\section{Background and Related Works}
\label{sec:background}
\subsection{Temporal Convolutional Networks}
\label{subsec:tcn} 
TCNs are a class of 1D-Convolutional Neural Networks (CNNs), whose peculiarity is in the use of \textit{causality} and \textit{dilation} in convolutional layers~\cite{bai2018empirical,lea2016temporal}. Causality constrains the convolution output $\mathbf{y}_{t}$ to depend only on inputs $\mathbf{x}_{\tilde{t}}$ with $\tilde{t} \leq t$, whereas dilation is a fixed gap $d$ inserted between input samples processed by the convolution, thus increasing its time receptive field without requiring more parameters.
A convolutional layer in a TCN implements the following function:
\vspace{-0.2cm}
\begin{equation}\label{eq:1d_conv}
\mathbf{y}_t^m = \sum_{i=0}^{K-1} \sum_{l=0}^{C_{in}-1} \mathbf{x}_{t-d\,i}^l \cdot \mathbf{W}_i^{l,m}
\vspace{-0.2cm}
\end{equation}
where $\mathbf{x}$ and $\mathbf{y}$ are the input and output feature maps, $t$ and $m$ the output time-step and channel respectively, $\mathbf{W}$ the filter weights, $C_{in}$ the number of input channels, $d$ the dilation factor, and $K$ the filter size.
In the original paper~\cite{bai2018empirical}, TCNs were proposed as fully-convolutional architectures, but modern embodiments also include other common layers such as pooling and linear ones~\cite{ren2020cloud, zanghieri2019robust}.
%

\subsection{State-of-the-art in PPG-based HR monitoring}
\label{subsec:HR_classical}
\begin{figure}
 \centering
\includegraphics[width=.8\columnwidth]{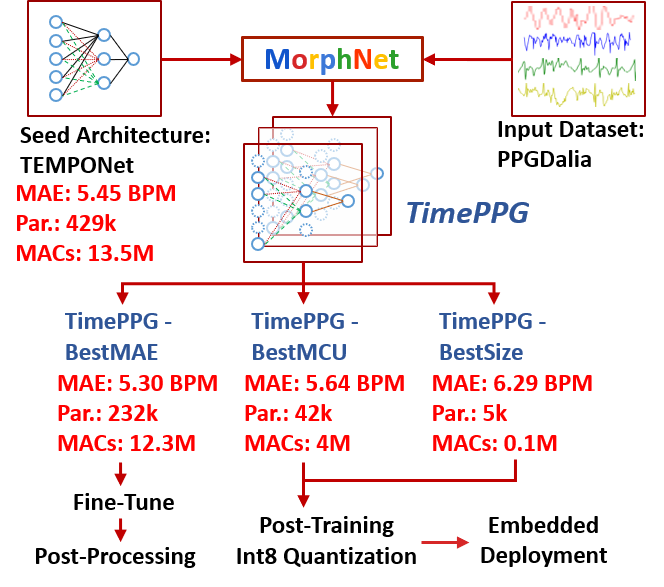}
  \vspace{-0.4cm}
  \caption{Proposed NAS and deployment flow. In red, the architectural parameters and MAE of the seed architecture and of three TimePPG Pareto points.}
  \label{fig:flow}
  \vspace{-0.4cm}
\end{figure}
HR monitoring through wrist-PPG is a relative new task, which both industry and researchers are exploring.
The seminal work of~\cite{troika2014} paved the way to algorithmic exploration in the field, releasing the first open-access dataset (the 12-subject SPC), and proposing a three-stage algorithm based on signal decomposition, spectrum estimation and spectral peak tracking (TROIKA), which achieves an average MAE of \unit[2.34]{BPM} BPM on SPC.
Later, \cite{joss2015} improved TROIKA using the spectral difference with the acceleration spectrum to clean the PPG signal from Motion Artifacts (MAs), reducing the error to \unit[1.28]{BPM}.
With the same goal, the authors of~\cite{mashhadi2015heart} applied Singular Value Decomposition to the acceleration data to extrapolate periodic MAs. Followed by some iterations of the Iterative Method with Adaptive Thresholding (IMAT), this method reduced the MAE to \unit[1.25]{BPM}.
Two approaches used Wiener filtering~\cite{temko2017accurate, chung2018finite} to clean PPG signals with accelerometer data, reaching a MAE of \unit[0.99]{BPM}.
Lastly, the complex five-steps pipeline called SpaMA~\cite{spama2016}, further reduced the MAE on this dataset to \unit[0.89]{BPM}.
More recently, \cite{huang2020robust} presented a time-domain algorithm which achieves \unit[4.6]{BPM} of MAE on the larger PPGDalia dataset with a 5-steps pipeline of light linear transformations, outperforming SpaMa~\cite{spama2016} and the CNN predented in ~\cite{DeepPPG2019} on these data, but performing worse on the smaller SPC. 
On the most challenging PPGDalia dataset, \cite{huang2020robust} is the SoA algorithm and thus used as a comparison in this work.

All aforementioned approaches share the common problem of including an high number of free hyper-parameters, 
which are not automatically learned, but hand-tuned to maximize performance, leading to an overfitting of the test dataset.
For instance, the authors of \cite{DeepPPG2019} have shown that the state-of-the-art results of~\cite{spama2016,schack2017computationally} on the SPC dataset, cannot be reproduced when cross-validation is used for optimizing hyper-parameters on the same dataset, leading to a  MAE increase of \unit[1.64-11.77]{BPM}.
Further, they saw a degradation of up to \unit[19.18]{BPM} MAE of classical approaches developed for SPC on the PPGDalia dataset, demonstrating that classical approaches hardly generalize over many subjects.

In recent years, researchers have tried to address this limitation with data-driven deep learning algorithms.
The works of \cite{cornet2019, DeepHeart2019} achieve comparable results to the ones of the classical methods, applying a CNN to frequency data and a CNN+LSTM (Long-Short Term Memory) to time data, respectively.
In \cite{DeepPPG2019}, the authors present a CNN architecture which outperforms two  classical methods~\cite{spama2016,schack2017computationally} on PPGDalia.
Despite these promising results, however, the deep learning architectures proposed in literature are still too complex to be embedded in a MCU-based wearable device.
For instance, the best architecture presented in \cite{DeepPPG2019}, based on a CNN ensembler, has 60M float32 parameters, while CorNet, a CNN+LSTM model, has 260k float32 parameters and requires 21.1 million operations.


%
\section{Time PPG}

Fig.~\ref{fig:flow} shows the complete flow proposed in this work.
As anticipated, we use an automatic tool to explore the space of possible TCN architectures for PPG-based HR monitoring. As a starting point for this exploration we use TEMPONet~\cite{zanghieri2019robust}, a TCN which shows impressive results in other bio-signals analysis tasks, originally developed for EMG-based gesture recognition.
TEMPONet includes a modular \textit{feature extractor}, composed of 3 convolutional blocks, each with two dilated convolutions, 1 strided convolution and 1 pooling layer.
The output channels in each block are 32, 64, and 128 respectively. 
The feature extractor is followed by a \textit{classifier} composed of 3 linear layers, contributing to almost half (200k) of the total network parameters. 
All layers use ReLU activations and batch normalization~\cite{ioffe2015batch}.
Further details on the network are omitted for sake of space, and can be found in~\cite{zanghieri2019robust}.

To adapt TEMPONet to our task, we modify the input layer to match the target dataset, described in detail in Section~\ref{sec:results}. The network takes as input raw sensor data generated by a PPG-sensor and a tri-axial accelerometer.
Samples are gathered at \unit[32]{Hz} and fed to the model in sliding windows of \unit[8]{s} with an overlap of \unit[75]{\%}, resulting in a 256$\times$4 input matrix.
We also remove the final classification layer, replacing it with a single neuron for regression, and use a \textit{LogCosh} loss for training.

\subsection{Design Space Exploration} 
\label{sect:arch_opt_mn}

%
\begin{figure}
  \centering
\includegraphics[width=0.9\columnwidth]{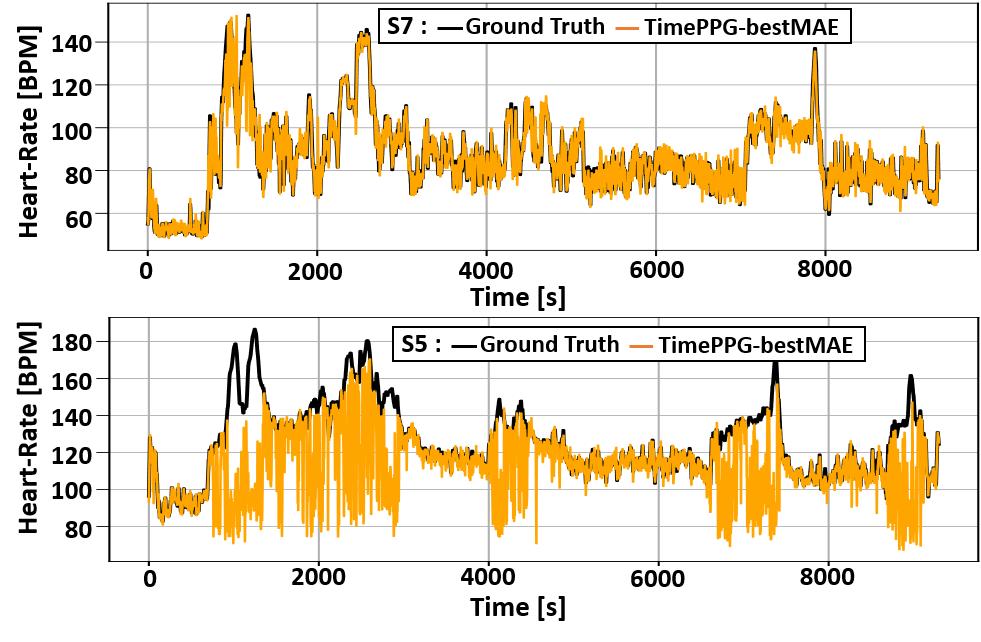}
  \vspace{-0.3cm}
  \caption{Ground truth and prediction of a well tracked patient (S7) and of the worst one (S5). Values above 140 BPM are not well estimated by our algorithm.}
  \label{fig:trace}
  \vspace{-0.4cm}
\end{figure}

NAS methods are the natural tools for automatically exploring novel NN architectures for a given task, acting on hyper-parameters such as the number and type of layers, the number of convolution filters, etc~\cite{tan2019mnasnet}.
%
%
However, standard NAS approaches require an enormous number of training iterations and are optimized for large-scale computer vision tasks, leading to oversized networks for simpler tasks.
Therefore, in this work we resort to MorphNet \cite{gordon2018morphnet} a new NAS algorithm which slightly reduces the search space using a seed network (TEMPONet in our case) in exchange for a dramatic reduction in complexity.

MorphNet limits the optimization to the \textit{number of channels} in each layer, and learns an optimal architecture which retains as much as possible the initial performance of the network, while reducing either its memory footprint, the number of required Multiply-and-Accumulate (MAC) operations, or a combination.
This is obtained with a two-step algorithm: 
first, the network size is reduced by applying group Lasso~\cite{yuan2006model}, a sparsifying regularizer which forces entire filter channels to small magnitude values. At the end of this phase, all channels whose magnitude is inferior to a threshold are eliminated.
Since compression is associated with an obvious performance penalty, MorphNet alternates it with an expansion step, in which the number of channels in all layers is uniformly up-scaled by a constant. 
Two regularizers are introduced in the original paper, to guide the search either on reducing the memory footprint (\emph{size regularizer}) or the number of MACs (\emph{flops regularizer}).
Importantly, the overall resources required by MorphNet are only slightly greater than those required to train the model once.
Further details are omitted for sake of space and can be found in~\cite{gordon2018morphnet}.

We explore the design space by applying MorphNet to TEMPONet, using grid search on the relative \textit{strengths} of the two regularizers and on the channels pruning threshold, thus
trading a penalty in MAE with a reduction in the number of parameters and complexity of the network.
Moreover, to further reduce the memory footprint for deploying our TCNs on resource-constrained MCUs, we also apply full-integer post-training quantization to the MorphNet outputs, converting them from float32 to int8~\cite{JahierPagliari2018a,Pagliari2018b}. 
As shown in Section~\ref{sec:results}, this simple approach yields a large set of Pareto-optimal TCNs, ranging from state-of-the-art accuracy to very small memory footprints.


%

\subsection{Post-Processing}
\label{sect:arch_opt_pp}

Despite being accurate on average, fully data-driven models such as TCNs can sometimes make large errors, especially when inputs deviate significantly from the distributions seen during training.
Fortunately, for HR monitoring, these errors can be easily filtered with a post-processing step,
%
%
which removes predictions that are not compatible with human physiology. Specifically, we impose a limit on the maximum relative HR variation over time. 
To this end, we compare the latest TCN prediction with the mean of the previous N predictions: if the difference between the two is larger than a threshold $th$, the predicted HR is \textit{clipped} to mean $\pm$ threshold.
We set N to 10 and $th$ to 10\% of the mean, identical for all patients.
%
%
\subsection{Fine-Tuning} 
\label{sect:arch_opt_ft}
Deep learning benefits from large amounts of training data, unfortunately not yet available in datasets for PPG-based HR monitoring, which include $<20$ patients. Therefore, subjects with particularly high/low HR are badly tracked by our TCNs, since their unique data distributions are not present elsewhere in the training set.
To underline this effect, Fig. \ref{fig:trace} showcases the accurate tracking on a patient with HR $<$ 140 BPM (S7), while miss-predicting intervals of time of S5 with HR $>$ 140 BPM (lower part of the figure).

We claim that this is indeed just an effect of the scarcity of data, and not a limitation of our models. 
To demonstrate it, in one of our experiments we apply an additional fine-tuning step to our trained TCNs.
Specifically, we fine-tune on the initial portion of data (25\%) relative to the patient under exam, with a low learning rate, freezing the weights of the first convolutional block. We then compute the MAE on the remaining 75\% of data.
Note that this step is hardly reproducible in the field, since collecting ground truth data for fine-tuning is hard. However, it mimics the effect of a larger dataset which would include data similar to those of a given test subject.
%
%

%
\section{Experimental Results}\label{sec:results}
We evaluate our TimePPG models on the PPGDalia dataset \cite{DeepPPG2019}, the largest publicly available dataset for PPG-based heart rate estimation. 
The dataset includes sensors' data from a PPG-sensor and a 3D-accelerometer, together with golden HR values from 15 subject and a total of 37.5 hours of recording.
We validate TimePPG models following the cross validation scheme proposed in \cite{DeepPPG2019}.
We train all TCNs with an Adam optimizer (learning rate = 1e-3, weight decay = 5e-4), and a batch size of 128 over 500 epochs, with an early stop mechanism with patience of 20.
%
%
All experiments are performed using Python 3.6, TensorFlow 1.14 \cite{tensorflow2015} and the Nucleo-STM32L476RG evaluation board, with \unit[128]{kB} of RAM, \unit[1]{MB} Flash and an average power consumption of \unit[12.1]{mW} at \unit[80]{MHz} \cite{st-L478}. 
\subsection{TimePPG Design Space Exploration Results} \label{sect:mn_res}
\begin{figure}
  \centering
\includegraphics[width=0.9\columnwidth]{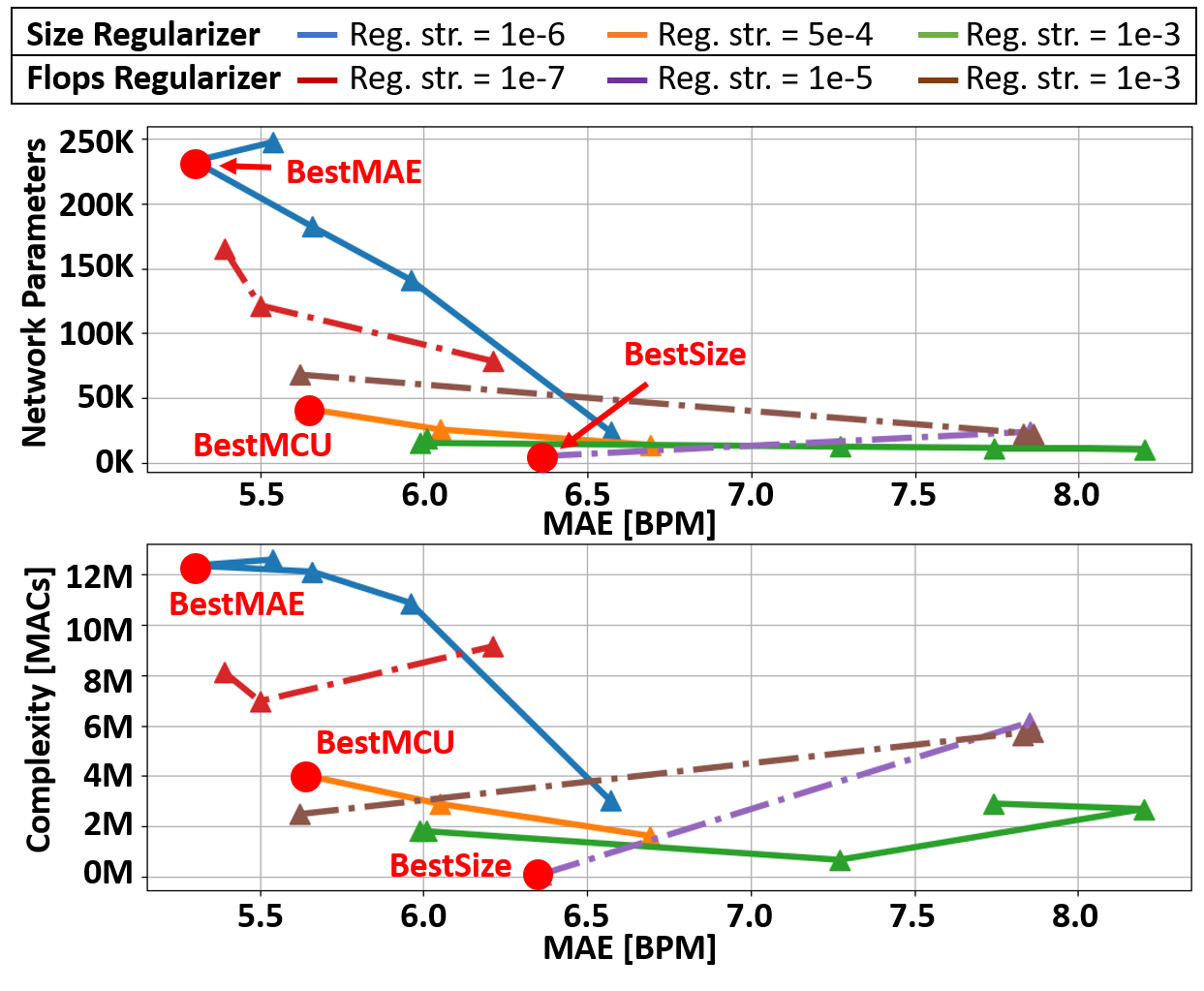}
  \vspace{-0.4cm}
  \caption{TimePPG results in the MAE vs. memory and MAE vs. MACs planes.}
  \label{fig:architecture}
  \vspace{-0.6cm}
\end{figure}

\begin{table*}
\centering
\caption{Comparison with state-of-the-art PPG-based HR monitoring algorithms. \textit{p}-values computed with non-parametric Mann-Whitney.}
\label{tab:SoA_comparison}
\vspace{-0.2cm}
\begin{tabular}{l|p{0.45cm}p{0.45cm}p{0.45cm}p{0.5cm}p{0.5cm}p{0.5cm}p{0.5cm}p{0.5cm}p{0.5cm}p{0.5cm}p{0.5cm}p{0.5cm}p{0.5cm}p{0.5cm}p{0.5cm}|c|c}
                        & S1   & S2 & S3   & S4            & S5            & S6            & S7            & S8            & S9            & S10           & S11           & S12           & S13           & S14           & S15           & Mean  & \textit{p}-value \\\hline\hline
Schack2017 \cite{schack2017computationally}          & 33.1         & 27.8         & 18.5         & 28.8         & 12.6         & 8.7          & 20.65         & 21.8         & 22.3         & 12.6         & 21.1         & 22.8         & 27.7         & 12.1         & 16.4         & 20.5          & $< 0.01$     \\ 
SpaMaPlus \cite{spama2016}           & 8.86          & 9.67          & 6.40          & 14.10         & 24.06         & 11.34         & 6.31          & 11.25         & 16.04         & 6.17          & 15.15         & 12.03         & 8.50          & 7.76          & 8.29          & 11.06          &  $< 0.01$    \\ 
STFT+CNN \cite{DeepPPG2019}    & 7.73          & 6.74          & 4.03          & 5.90          & 18.51         & 12.88         & 3.91          & 10.87         & 8.79          & 4.03          & 9.22          & 9.35          & 4.29          & 4.37          & 4.17          & 7.65           &  $< 0.01$    \\ 
TAPIR \cite{huang2020robust}    & 4.50          & 4.50          & 3.20          & 6.00          & 5.00          & 3.40          & 2.80          & 6.30          & 8.00          & 2.90 & 5.10          & 4.70          & 3.10          & 5.00          & 4.10          & 4.57          &     $< 0.01$  \\ 
CurToSS \cite{zhou2020heart}    & 5.40          & 4.30          & 3.00          & 8.00          & \textbf{2.20} & \textbf{2.80} & 3.30          & 8.50          & 12.60         & 3.60          & \textbf{3.60} & 6.10          & 3.00          & 5.50          & 3.70          & 5.04          &     $0.02$    \\ \hline\hline
TimePPG-BestMAE             & 4.51          & 3.37          & 2.33          & 5.25          & 14.68         & 4.76          & 2.37          & 8.04          & 8.75         & 3.3          & 5.19          & 8.08          & \textbf{2.29} & 3.02          & \textbf{3.49}          & 5.30          &   n.a.     \\ 
+ Post-Processing       & 4.01          & 3.16          & \textbf{2.27} & 4.62          & 14.96         & 4.28          & 2.58          & 6.02          & 7.61          & \textbf{2.89}          & 4.79          & 6.95          & 2.54          & \textbf{3.01} & 3.56 & 4.88          &  n.a.      \\ 
+ Fine Tuning & \textbf{3.17} & \textbf{2.74} & 3.13          & \textbf{4.25} & 4.88         & 3.7          & \textbf{2.48} & \textbf{5.19} & \textbf{7.00} & 3.47         & 3.67          & \textbf{3.91} & 2.85        & 3.55          & 3.6         & \textbf{3.84} &  n.a.   \\ \hline
\end{tabular}
 \vspace{-0.5cm}
\end{table*}

Fig.~\ref{fig:architecture} shows some of the models obtained applying MorphNet to the seed network with different regularizer strengths and pruning thresholds.
Only models achieving a MAE lower than 7 BPM are reported.
In this experiment, MAE results refer solely to the TCN output, without post processing or fine tuning.

As shown, our design space exploration spans more than one order of magnitude both in terms of TCN parameters (5k-230k) and  MACs (0.1M-12M).
Three relevant Pareto-optimal models are higlighted in the figure, called BestMAE, BestMCU, and BestSize.
The former is the one achieving the lowest overall MAE (5.3 BPM) and is analyzed in detail in Section~\ref{sect:soa}. It requires around 230k parameters and 12M MACs.
BestMCU is the best performing model that meets the memory constraints imposed by the target MCU; it requires 41.7k parameters, and 4M MACs and achieves a MAE of 5.64 BPM, i.e. a model compression of 5.6$\times$ w.r.t. BestMAE, with a MAE increase of just 0.34 BPM.
BestSize is the smallest model found in our design space exploration, obtained using MorphNet's flops regularizer with a strength of 1e-5 and a pruning threshold of 0.01. The model has just 5.09k parameters (46$\times$ compression) and less than 100k MACs, with a MAE of 6.29 BPM (0.99 increase). 
Section~\ref{sect:emb_res} analyzes the execution metrics of the latter two TCNs on the STM32L476RG.

\subsection{TimePPG-BestMAE: state-of-the-art comparison} \label{sect:soa}

\begin{table}
\centering
\begin{threeparttable}
\caption{Porting of different models of our TimePPG on an STM32L4 with 128 kB RAM memory and 1 MB Flash memory. }
\label{tab:micro_deploy}
\begin{tabular}{p{1.5cm}|c|c|c|c}
Model                 & Ram/Flash [kB] & E. [mJ] & Time [ms] & MAE [BPM] \\\hline\hline
BestMCU            &  84.0 / 160.0      &   5.17      &   427.0   &     5.64      \\
BestMCU\tnote{\textdagger} & 24.9 / 94.7      &     4.44    &   367.0   &    7.06   \\
BestSize            & 11.5 / 16.5       &     0.21    &   17.1   &    6.29   \\
BestSize\tnote{\textdagger} & 7.21 /   8.07     &    0.23     &   19.0   &   7.55   \\\hline
\end{tabular}
\begin{tablenotes}
\item [\textdagger] With post-training int8 quantization. Dilation reduced to one, with bigger filters to maintain the receptive field, to cope with toolchain limitations.
\end{tablenotes}
\end{threeparttable}
\vspace{-0.3cm}
\end{table}
Table~\ref{tab:SoA_comparison} compares our proposed BestMAE TCN with different state of the art methods, including both classical and deep learning approaches. 
We report the results of the TCN alone, as well as those obtained after the application of our proposed post-processing and with fine-tuning.

TimePPG outperforms previous deep learning approaches, such as the CNN from \cite{DeepPPG2019} and the Generative Adversarial Network (GAN)-based method reported in \cite{sarkar2020cardiogan} (not in the table since individual subjects performance are not reported in the original paper), achieving a mean MAE of 5.3 BPM vs. 7.65 BPM (CNN) and 8.3 BPM (GAN).
Further, our model has 230k parameters, i.e. 260$\times$ smaller than the CNN ensemble of \cite{DeepPPG2019}.

Among classical algorithms, Schack2017 and SpaMaPlus are optimised on a different dataset (IEEE training \cite{troika2014}), while CurToSS and TAPIR are tailored to PPGDalia.
Note that while the four methods employ similar filtering and peak tracking algorithms, the ones optimized for PPGDalia substantially outperform the others, demonstrating the intrinsic overfitting due to parameter tuning in classical methods.
%
With post-processing, TimePPG-BestMAE is comparable to CurToSS (-0.16 BPM) and TAPIR (+0.31 BPM).
In particular, while our method achieves comparable performance on the majority of the patients, it is substantially outperformed by classical methods on subject 5 (MAE of 2 and 5 BPM vs 14.96 BPM of TimePPG with post processing), which has HR values in the range 160-180 BPM. 
Data-driven deep learning methods are less accurate in predicting such large HR values, since no sample in that range is present in the dataset used for training models validated on subject 5.
This is confirmed by the fact that also the CNN from~\cite{DeepPPG2019} similarly fails in estimating the HR of this subject, reaching a MAE of 18.51 BPM.

To mimic a wider training dataset including all realistic HR ranges, we applied the fine-tuning described in Section~\ref{sect:arch_opt_ft}.
As shown in Table~\ref{tab:SoA_comparison} the fine-tuned TimePPG-BestMAE obtains the lowest mean MAE of 3.84 BPM, outperforming all state-of-the-art methods.
Note that this improvement is mainly due to subject 5, whose MAE decreases from 14.68 to 4.88 BPM.

\subsection{Embedded Deployment Results} \label{sect:emb_res}
Table~\ref{tab:micro_deploy} summarizes the results obtained deploying the previously described BestSize and BestMCU TCNs on the STM32L476RG. 
Specifically, we deploy both float32 and int8-quantized variants of each TCN.
Unfortunately, the toolchain offered by STM to deploy neural network models on their MCUs, called X-CUBE-AI \cite{CubeAI}, does not support int8 quantization with dilation factors higher than 1. To cope with this limitation, we are obliged to use larger filters of size $d\times(K-1)+1$ and dilation 1 to maintain the original receptive field of the float models. This actually increases the number of parameters in the networks, resulting in a more complex network -- 1.8$\times$ higher number of parameters and 2.7$\times$ more computation.
Overall, the table shows that latency and energy consumption metrics obtained with our methodology starting from a single seed model span more than one order of magnitude. On one extreme, the float version of BestMCU achieves a mean MAE of 5.64 BPM with an energy consumption of \unit[5.17]{mJ} and \unit[427]{ms} of latency.
Alternatively, we can use BestSize to lower the energy consumption to just \unit[0.21]{mJ} and the latency to \unit[17.1]{ms} at the cost of a higher MAE, 6.29 BPM.

Note that these results could be improved by \textit{i)} applying a quantization-aware training and \textit{ii) } manually porting the network to the MCU implementing $d > 1$ with int8 filters. These steps would be objects of our future work.
Despite these limitations, BestSize\tnote{\textdagger} model result in a Flash occupation of just 8.07 KB, which could be very promising for the porting on commercial ultra-low-power wrist-worn devices, but with almost equal computation compared to the float32 model.

\section{Conclusions}
The efficient execution of HR monitoring algorithms is a critical enabler for the personalized health care.
In the direction of employing lightweight algorithms for HR, we have proposed a set of TCN regressors, all automatically derived from a single seed model using NAS.
With our exploration, we spanned a wide range in all metrics, reaching as low as 3.84 BPM average MAE,  \unit[0.21]{mJ} of energy per inference and 8.07 KB of memory footprint, enabling the deployment of our models even to very tiny MCUs with small embedded flash memories.

\footnotesize
\bibliographystyle{IEEEtran}
\bibliography{references}

\end{document}